\documentclass{aa}
\usepackage{txfonts}
\usepackage{graphics}
\usepackage{epsfig}
\usepackage{graphicx}	
\usepackage{amssymb}
\usepackage{afterpage}
\usepackage{natbib}
\begin{document}

\def\deg{$^{\rm o}$}
\def\ico{I$_{\rm CO}$}
\def\i14{I$_{\rm 1.4}$}
\def\Qco{Q$_{\rm CO/RC}$}
\def\qco{q$_{\rm CO/RC}$}
\def\qfir{q$_{\rm FIR/RC}$}
\def\matteo{Paper I}
\def\P{$\bar{\rm P}$}
\def\n0{N$_{0}$}
\def\tc{$t_{\rm c}$}
\def\tsyn{$t_{\rm syn}$}

\defcitealias{matteo05}{Paper~I}

\title{Radio spectral index images of the spiral galaxies NGC 0628, NGC 3627, and NGC 7331}

\author{ R. Paladino \inst{1,2} \and M. Murgia \inst{2,3} \and E. Orr{\`u} \inst{1}}
\offprints{R. Paladino, rpaladin@ca.astro.it}

\institute{
Institut fur Astro- und Teilchenphysik Universitat Innsbruck, Technikerstrasse 25, 6020 Innsbruck, Austria
\and
INAF\,-\,Osservatorio Astronomico di Cagliari, Loc. Poggio dei Pini, Strada 54,I-09012 Capoterra (CA), Italy
\and
Istituto di Radioastronomia - INAF, Via Gobetti 101, I-40129 Bologna, Italy
}

\date{Received; Accepted}

\abstract{}{In order to understand the cosmic ray propagation mechanism in galaxies, 
and its correlation with the sites of star formation, we compare the spatially resolved 
radio spectral index of three spiral galaxies with their IR distribution.}
{We present new low-frequency radio continuum observations 
of the galaxies NGC 0628, NGC 3627, and NGC 7331, taken at 327 MHz with 
the Very Large Array.}
{We complemented our data set with sensitive archival observations at 1.4 
GHz and we studied the variations of the radio spectral
index within the disks of these spiral galaxies. We also compared 
the spectral index distribution and the IR distribution, using 70 $\mu$m  
Spitzer observations.}
{We found that in these galaxies the non-thermal spectral index is anticorrelated with 
the radio brightness. Bright regions, like the bar in NGC 3627 or 
the circumnuclear region in NGC 7331, are characterized by a flatter spectrum 
with respect to the underlying disk. Therefore, a systematic steepening 
of the spectral index with the increasing distance from the center of these 
galaxies is observed. Furthermore,
by comparing the radio images with the 70 $\mu$m images of the Spitzer 
satellite we found that a similar anticorrelation exists
between the radio spectral index and the infrared brightness, as expected 
on the basis of the local correlation between the radio continuum and the 
infrared emission. 
Our results support the idea that in regions of intense star formation the 
electron diffusion must be efficient. The observed  
anticorrelation between radio brightness and spectral index, may imply that the cosmic ray density  
and the magnetic field strength are significantly higher in these regions than in their 
surroundings.}

\maketitle

\begin{table*}
\caption{Summary of observations.}
\begin{center}
{\bf A.} Sources parameters\\
\smallskip
\begin{tabular}{ccccccc}
\hline
\noalign{\smallskip}
Name & $\alpha$(J2000) & $\delta$(J2000)& d & i& S$_{\rm RC}$&S$_{\rm FIR}$ \\
&($h\  m\  s\ $)   & (\degr\ \arcmin\ \arcsec\ )& (Mpc)&(\degr)&(mJy)& (Jy) \\
\noalign{\smallskip}
NGC\,0628& 01 36 41.7 &+15 46 59&7.3&24&180&20.86\\
NGC\,3627& 11 20 15.0& +12 59 30& 11.1&63     & 458  & 67.8  \\
NGC\,7331&22 37 04.1& +34 24 56&15.1 &62&372   &35.29  \\  
\noalign{\smallskip}
\hline
\end{tabular}
\smallskip

{\bf B.} Observations details \\
\smallskip
\begin{tabular}{ccccccccc}
\hline
\noalign{\smallskip}
Name & IF & Frequency & Bandwidth& Configuration&Visibilities& HPBW& rms&Date \\
& Number& MHz& MHz&&Number&$\arcsec \times \arcsec$&mJy/beam &\\
\noalign{\smallskip}
NGC\,0628&2&321.5-327.5 &3.125 &C&524446& 65 $\times$ 57 &2.3&01 sep 05\\
NGC\,3627&2&329.1-323.1 &6.250 &B&662708& 20 $\times$ 16.6 &2    & 30 oct 07         \\         
NGC\,7331&2&329.1-323.1 &6.250 &B&349475&17.5 $\times$ 15.5&1.1 & 03 nov 07      \\


\noalign{\smallskip}
\hline
\end{tabular}

\end{center}
\begin{list}{}{}
\item[]
{\bf A.} Coordinates, distances, inclinations from \cite{helfer03}. Inclination is 
defined assuming i=0\degr\ for face-on galaxies. The  S$_{\rm RC}$ of NGC\,0628, 
and NGC\,3627 are 1.4 GHz global flux from \citealt{condon90}, the  
S$_{\rm RC}$ of NGC\,7331 is from \cite{condon87}.
S$_{\rm FIR}$ is 60 $\mu$m flux, taken from \citealt{rice90} for NGC\,0628, and 
for NGC\,7331, while from the IRAS Bright Galaxy Sample \citep{soifer89} for 
NGC\,3627. 

{\bf B.} VLA observational properties. 
\end{list}

\label{tab1}

\end{table*}

\section{Introduction}

It is well known that the radio continuum of spiral galaxies consists of 
a mixture of non-thermal and thermal 
emission \citep[e.g.][]{condon92}. The non-thermal component is due 
to the synchrotron
emission of cosmic-ray (CRe) electrons with GeV energies spiralling in the 
galaxy's magnetic field.
The thermal one is due to the free-free emission from ionized gas 
at $T\sim 10^4$ K.
The two emissions are characterized by different spectra. The synchrotron 
component generally has a steep
spectral index $\alpha\sim 0.8$ ($S_{\nu}\propto \nu^{-\alpha}$) and dominates 
at low-frequencies
 while the free-free emission has a much flatter spectrum,
 $\alpha\simeq 0.1$, and may become relevant
 at frequencies greater than about 10 GHz \citep{niklas97}.
Synchrotron emission is thus best studied at frequencies below a few GHz.
At these frequencies  
close correlations are observed between the radio continuum and several 
star formation tracers like
the $H_{\alpha}$ emission from ionized gas, the infrared radiation (IR) 
from thermal dust \citep[e.g.][]{yun01}, and the CO emission 
from molecular clouds (\citeauthor{rickard77} \citeyear{rickard77},
\citeauthor{israel97} \citeyear{israel97}, \citeauthor{adler91} 
\citeyear{adler91}, \citeauthor{murgia02} \citeyear{murgia02}). 
It has been established that these correlations hold not only  
on global but also on local scales, down to hundreds of parsecs, 
within the disks of spiral galaxies (\citeauthor{matteo05} \citeyear{matteo05}, 
\citeauthor{io06} \citeyear{io06}).
Although it is commonly accepted that all these correlations ultimately 
originate from the process
 of massive stars formation itself, the details of the physical 
mechanisms on how this occurs
 are not completely understood.

One main difficulty in the study of the non-thermal radio continuum 
is due to the fact that 
the synchrotron radiation essentially traces the product of CRe and 
magnetic field energy densities.
Disentangling these two contributions is not directly possible unless 
making hardly verifiable assumptions
(like e.g. energy equipartition between field and particles).
A second source of uncertainty is related to the diffusion length of 
CRe electrons which is poorly 
known. The CRe diffuse from their injection (or re-acceleration) sites 
and, as a consequence,  
their density and energy spectrum at a given location in the disk is 
the result of the balance between the efficiencies of different processes 
such as the 
 injection (or re-acceleration) rate, the confinement time, and the energy 
losses.

Since the mean free path of dust-heating photons ($\sim$100 pc) is far 
shorter than the presumed diffusion length of CR electrons \citep{bicay89}, the first 
attempts to model the 
propagation of CRe have been made by comparing radio and IR images of 
spiral galaxies.
\cite{bicay90} smeared IRAS images of spiral galaxies by using 
parametrized kernels in order to
match the observed morphology seen in the radio. They found that 
the diffusion length of CRe electrons is of the 
order of $\sim$ 1-2 kpc. Recently \cite{murphy06L} applied this 
phenomenological image-smearing model to 
new {\it Spitzer} images. They found that the galaxies examinated exhibit a 
wide range of ages and spreading scale lengths of 
their CR electrons, going from the youngest 
populations, recently injected within star-forming 
regions (l$\sim$500 pc), to the oldest CR electrons, making 
up a galaxy's underlying synchrotron disk (l $\sim$ 3 kpc). 

The way we will use to understand the cosmic ray diffusion 
mechanism, and thus its correlation with the sites of intense star formation,
is the comparison of spatially resolved radio spectral index images of galaxies
and their IR distribution.
Such a study is only now possible due to the unprecedented high spatial 
resolution and sensitivity 
of the {\it Spitzer Space Telescope} at IR wavelengths.
We recently studied the point-by-point relation between the 
radio spectral index and the IR emission in the galaxy M51 
(Paladino et al., 2006). By comparing the {\it Spitzer} 70 $\mu$m image and 
the spectral index image obtained between 1.4 and 4.9 GHz, we found 
an anticorrelation between the IR emission and the spectral index.
Regions in which the IR is higher than the average tend to have
 a flatter radio spectrum with respect to the surroundings. 
This result suggests that
the CRe electrons confinement time in the star forming regions is shorter 
than their radiative lifetime
and that they age as they diffuse in the disk (Murgia et al 2005).

\begin{figure*}[p]
\begin{center}
\includegraphics[width=0.56\textwidth]{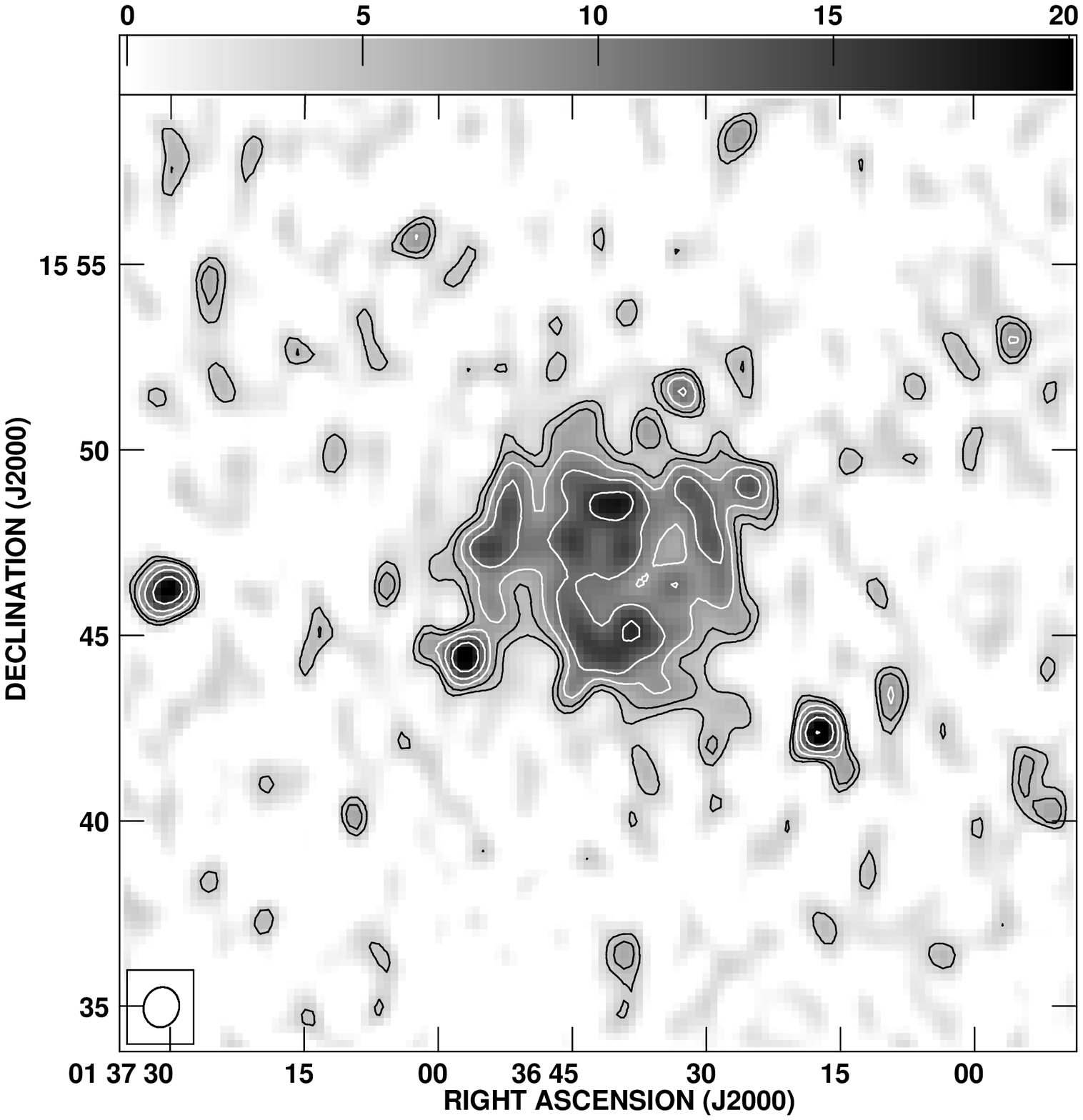}
\end{center}
\vspace{-1cm}
\caption[]{{\bf{Galaxy NGC\,0628}}: Radio image  
at 327 MHz. The FWHM is 65.4\arcsec$\times$~57\arcsec. 
Contour levels start from a level of 2 $\sigma$ (i.e. 4 mJy/beam) 
and scale by a factor of $\sqrt{2}$. Grey scale flux range is 
0 -- 20 mJy/beam. }
\label{N0628_90cm}

\begin{center}
\includegraphics[width=0.58\textwidth]{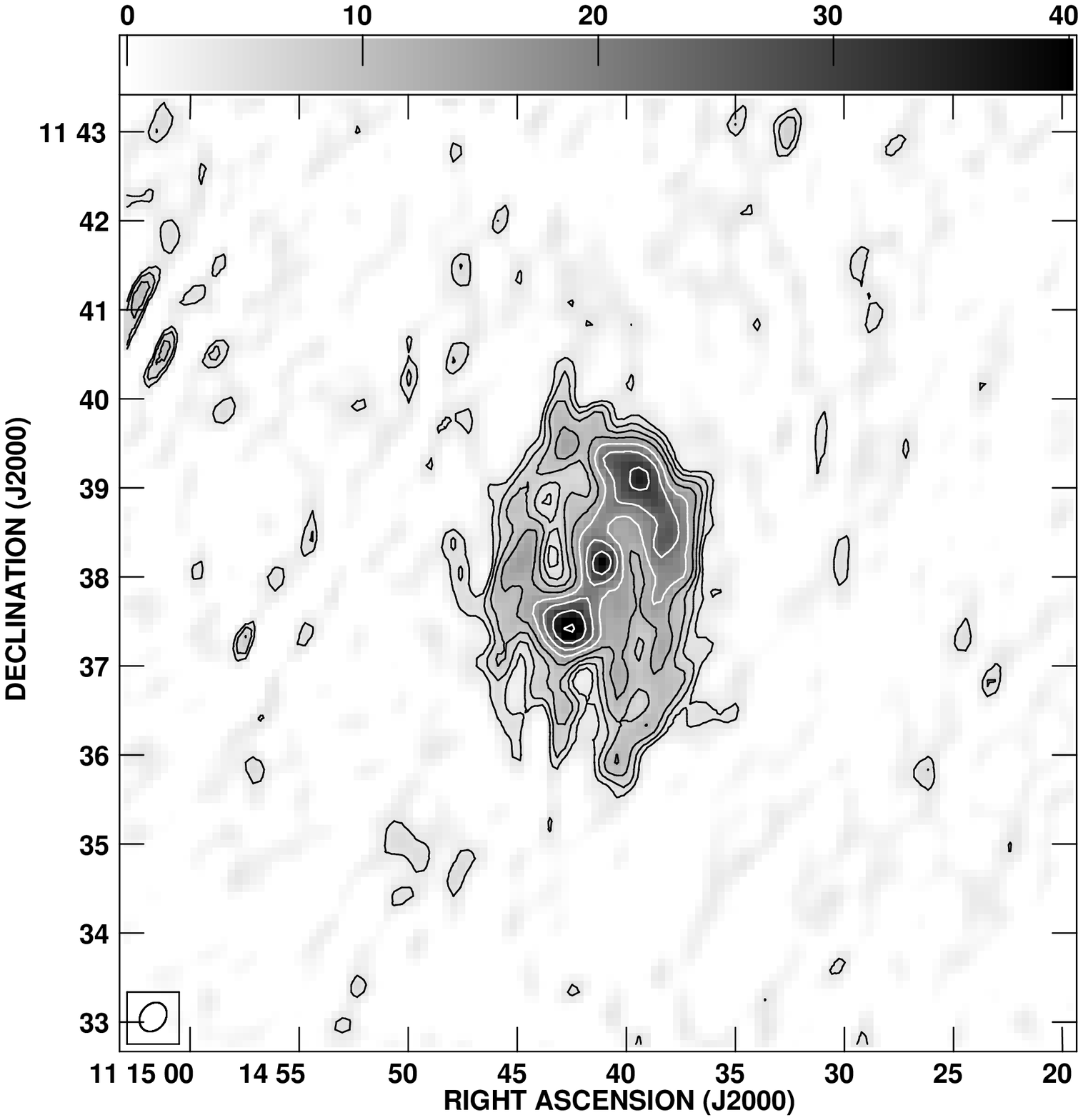}
\end{center}
\vspace{-1cm}
\caption[]{{\bf{Galaxy NGC\,3627}}: Radio image
at 327 MHz. The FWHM is 21.0\arcsec$\times$~16.6\arcsec.
Contour levels start from a level of 2 $\sigma$ (i.e. 4 mJy/beam)
and scale by a factor of $\sqrt{2}$. Grey scale flux range is
0 -- 40 mJy/beam.}
\label{N3627_90cm}
\end{figure*}

\begin{figure*}
\begin{center}
\includegraphics[width=0.58\textwidth]{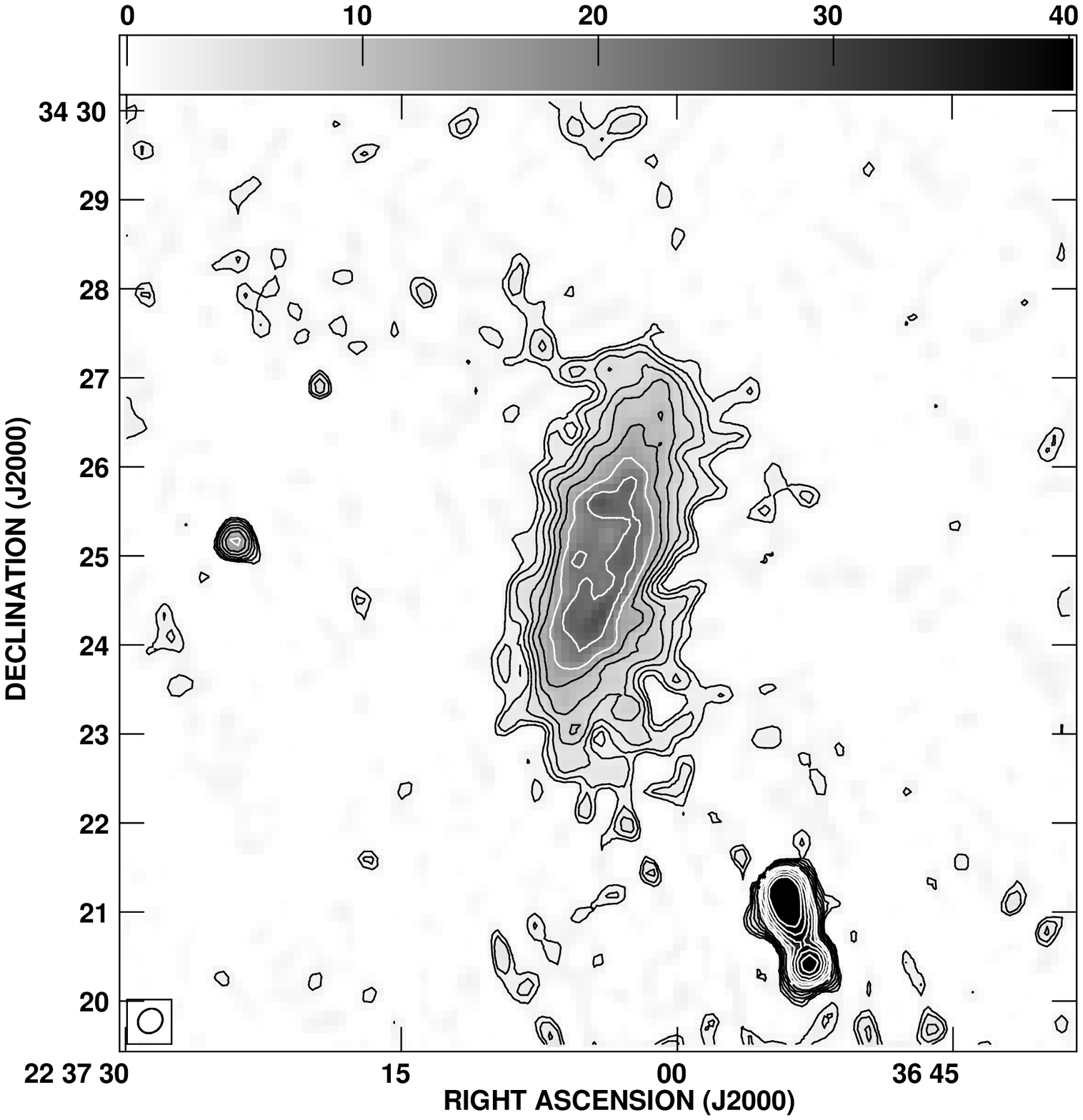}
\end{center}
\vspace{-1cm}
\caption[]{{\bf{Galaxy NGC\,7331}}: Radio image
at 327 MHz. The FWHM is 17.5\arcsec$\times$~15.5\arcsec.
Contour levels start from a level of 2 $\sigma$ (i.e. 2 mJy/beam)
and scale by a factor of $\sqrt{2}$. Grey scale flux range is
0 -- 40 mJy/beam.}
\label{N7331_90cm}
\end{figure*}

This work would be a starting point for the program of extending this 
kind of analysis 
to a large sample of 
galaxies. Here we report on the spectral index analysis of three 
spiral galaxies:
 NGC\,0628, NGC\,3627, and NGC\,7331.  We observed these galaxies
 at 327 MHz with the Very Large Array (VLA).
In order to obtain detailed spectral index maps 
we complemented our data set with sensitive archival observations 
at 1.4 GHz.
Since at these frequencies the contribution from thermal 
emission is likely to be minimized the spectral index can be 
used to study the spectrum of the non-thermal emission.

The paper is organized as follows: in 
Sect. \ref{obs}  we describe the observations and 
the data reduction.
The spectral index analysis and the corresponding results are 
reported in Sect. \ref{res}.
Finally, we summarize our results in Sect. \ref{summ}.

\section{Observations and data reduction}
\label{obs}

\subsection{VLA 327\,MHz observations}
\label{images}

Our aim is to investigate the variations of the non-thermal
spectral index in the sample of 22 nearby spiral galaxies presented in
Murgia et al. (2005) and Paladino et al. (2006), by mean of low-frequency
radio observations. In this work we report on the first results for three
galaxies taken from the aforementioned sample.

We present new radio observations of 
the spiral galaxies NGC\,0628, NGC\,3627, and NGC\,7331 at 327 MHz 
with the VLA. 
The basic parameters of these galaxies are listed in Table \ref{tab1}.A. 
The galaxy NGC\,0628 is a clear grand-design spiral galaxy seen almost 
face-on. In the UV bright knots embedded in diffuse emission trace
 the spiral pattern and many of these knots are also bright in 
$\rm H_\alpha$ \citep{marcum01}. Futhermore these authors have 
suggested that the entire disk has undergone active star formation 
within the past ~500 Myr and that the
inner regions have experienced more rapidly declining star 
formation than the outer regions.
NGC\,3627, the brightest member of the Leo Triplet, is 
an interacting spiral galaxy with a bar and two asymmetric spiral arms. 
The western arm is accompanied by weak traces of star formation, while 
the eastern arm contains a straight actively star-forming segment in its 
inner part \citep{soida01}.
NGC\,7331 is a relatively highly inclined spiral galaxy with a ring, visible 
in CO \citep[e.g.][]{regan01}, radio continnum \citep[e.g.][]{cowan94}, 
submillimeter \citep{bianchi98} and {\it Spitzer}-infrared 
(\citeauthor{regan04} \citeyear{regan04},
\citeauthor{regan06} \citeyear{regan06}) images.

We observed NGC\,3627 and NGC\,7331 with the VLA in B configuration while
 NGC\,0628 was observed with the VLA in C configuration. 
A summary of the observations, including the VLA 
configuration, frequency, bandwidth, visibilities number, date of observations, 
is reported in Table \ref{tab1}.B. 

The observations were conducted in spectral line mode 
(see Tab. \ref{tab1}.B for  details on frequencies and bandwidths),  
in order to reduce the effects of bandwidth smearing  and to allow 
for more accurate 
radio-frequency interference (RFI) excision.

The data were calibrated and imaged with the NRAO package AIPS (Astronomical 
Image Processing System).
Both flux density and bandpass calibration were based directly on observations 
of the strong calibrator sources 3C48, 3C123, 3C147, 
and 3C286. The amplitude gains were tied to the flux scale 
of \cite{baars77}.
For the initial phase calibration we used, when available, further 
calibrators closest to the observed sources.
A careful data editing operation has been made in order to remove 
RFI present in individual channels. At the end of 
this procedure $\sim$ 10 $\%$ of data were flagged in the observations
of NGC\,0628. A higher fraction of data, about 35 $\%$ and 28$\%$ for 
NGC\,3627 and NGC\,7331, respectively, was flagged in the 2007 runs, 
when some EVLA antennas 
started to be included in the VLA.
 
The visibility data have been spectrally averaged, retaining 5 channels 
with a resolution 
of 488 KHz for NGC\,0628, and 6 channels with a resolution of 781 KHz 
for NGC\,3627 and NGC\,7331.

The entire field was imaged using the AIPS program IMAGR in 3D mode.
Owing to its large field of view at 327 MHz, the VLA is not coplanar at
 this frequency.
Therefore, the primary beam was divided in small overlapping ``facets'' 
over which the VLA can be 
safely assumed to be coplanar \citep{cornwelper92}. Confusion lobes of 
surrounding sources 
far from the center of the field are often so strong that their 
side-lobes contribute measurably to the noise level in the central region. 
Thus, by using  the NRAO VLA Sky Survey (NVSS)
catalog as a reference, we also cleaned the strongest sources over an area
of about $\sim~60\degr$ in radius by placing small facets around each of 
them (task SETFC).

Final images were created by an iterative process of CLEANing and 
self-calibration including all these ``facets''. In the self-calibration 
process, 
we used only 
the strongest components inside the primary beam in order
to increase the accuracy of the phase calibration model. 
The use of smaller clean boxes 
decreased the possibility of ``over-cleaning'' and minimized the CLEAN bias 
\citep [see][ for details]{condon98}.

Figures \ref{N0628_90cm}-\ref{N7331_90cm} show 
the 327 MHz images of the three observed sources, with 
typical noise level reported in Table \ref{tab1}.B.

NGC\,0628, which has the largest angular extent ($\sim 6\arcmin$) 
among the three galaxies, has been observed with the VLA in C configuration
with an angular resolution of 65\arcsec$\times$57\arcsec. 

The observations of NGC\,3627 and NGC\,7331 ($\sim$ 5\arcmin) were taken with the 
VLA in B configuration at a resolution of 20\arcsec$\times$16.6\arcsec 
and 17.5\arcsec$\times$15.5\arcsec, respectively. 

\subsection{Archival VLA observations at 1.4 GHz}
\label{20cm}

For the purposes of the spectral index imaging,
we complemented our data set with archival observations at 1.4 GHz 
with the VLA in
 D and C configurations. The archival data we recovered is described 
in Table \ref{20cmtab}.
The data were calibrated and imaged with the package AIPS following 
the standard procedures. The coverage of the spatial frequencies of 
the 1.4 GHz 
observations is very similar to that at 327 MHz. 
However, in order to ensure a perfect match in angular resolution, we 
convolved the final images at both frequencies to the largest beam 
(see section \ref{local_spix}).

\begin{table}[h!]
\caption{VLA archival observations at 1.4 GHz.}
\begin{center}
\begin{tabular}{ccccc}
\hline
\hline
\noalign{\smallskip}
Source& VLA&HPBW&rms&  Program ID\\
& config&$\arcsec \times \arcsec$&$\mu$Jy/beam&\\
\noalign{\smallskip}
\hline
\noalign{\smallskip}
NGC\,0628& D& 42.72 $\times$ 40.06 &54 &AA161\\
NGC\,3627& C&16.52 $\times$  14.02 &150 &AS541\\
NGC\,7331& C&17.49 $\times$ 15.48 &40 &AB345 \\
\hline
\hline
\end{tabular}
\end{center}
\label{20cmtab}
\end{table}

\subsection{IR {\it Spitzer} images}
The galaxies studied in this work were observed as part of the {\it Spitzer} 
Infrared Nearby Galaxies Survey (SINGS; \cite{sings03}) and IR images 
have been released. For details on observations and data processing see 
delivery document for Fifth (and last) SINGS Data Delivery, April 
2007\footnote{http:$//$data.spitzer.caltech.edu$/$popular$/$sings$/$20070410\_enhanced\\\_v1$/$Documents$/$sings\_fifth\_delivery\_v2.pdf}.
  
The observations obtained with the Multiband Imaging Photometer for 
{\it Spitzer} (MIPS; \cite{rieke04}) at 70 $\mu$m have resolution of 
18\arcsec. We convolved these images to the beam of our 327 MHz radio images in 
order to compare the spectral index distribution with the IR one.
As reported in the Data Release, due to special processing,   
the quality of  the NGC\,7331 70 $\mu$m 
data is somewhat worse than the quality of the other 70 $\mu$m 
images released. 
However, we use this image to compare the spectral index and IR 
distribution in NGC\,7331, handling with care 
the results obtained from this analysis. 

\section{Analysis and results}
\label{res}

\subsection{Integrated radio spectra}
\label{intspec}

\begin{table}[h!]
\caption{Integrated flux densities on the observed sources.}
\begin{center}
\begin{tabular}{cccc}
\hline
\hline
\noalign{\smallskip}
Source& Frequency& Flux density&Reference\\
& MHz& Jy&\\
\noalign{\smallskip}
\hline
\noalign{\smallskip}
NGC\,0628& 57.5& 2$\pm$1&1\\
&324.4   &  0.49	$\pm$0.03  & This work\\ 
&1400	&  0.15	$\pm$0.01 & 2\\
&1515 & 0.16 $\pm$ 0.01 & This work\\
&4850	&  0.060 $\pm$	0.005&3\\

\noalign{\smallskip}
\hline
NGC\,3627 & 57.5&2.3$\pm$ 0.7&1	\\
&160 &1.6 $\pm$	0.3$^a$& 4\\
&325.5 & 0.97 $\pm$ 0.03& This work\\
&408&0.93 $\pm$	0.05$^a$& 5\\
&611& 	     0.87 $\pm$ 0.1$^a$& 4\\
&1370&	  0.5  $\pm$   0.01& 6\\
&1425 & 0.38  $\pm$   0.01& This work\\
&2695&      0.2  $\pm$      0.04 & 4\\
&4850&	  0.14$\pm$   0.02& 7\\
\noalign{\smallskip}
\hline
NGC\,7331 & 74&1.12$\pm$  0.18& 8	\\
& 325 &  0.935$\pm$   	0.004&9 \\
& 325.9& 0.94$\pm$ 0.02& This work\\
&1489   &   0.32$\pm$ 	0.01& This work\\
&1400 & 0.33 $\pm$ 0.01 & 2\\
&1369 & 0.35 $\pm$ 0.01 & 6\\
&2380	&  0.18$\pm$	0.01&	10\\
&4850	&  0.096$\pm$	0.013&	7\\
\hline
\hline
\end{tabular}
\end{center}
\label{sed}
\begin{list}{}{}
\item[]
$^a$ Flux density scale from \cite{wyllie69}.
\item[]
References and telescopes: 
\begin{enumerate}
\item \cite{israel90}: Clark Lake radio Telescope 
\item \cite{NVSS}: NVSS -- NRAO VLA D array
\item \cite{GB6}: NRAO Green Bank Telescope 
\item \cite{CSIRO75}: CSIRO telescope
\item \cite{large81}: Molonglo Radio Telescope
\item \cite{braun07}: Westerbork Telescope WSRT 
\item \cite{gregory91}: NRAO Green Bank
\item \cite{cohen07}: NRAO VLA B array
\item \cite{rengelink97}: WENSS -- WSRT
\item \cite{dressel78}: Arecibo Telescope.
\end{enumerate}
\end{list}
\end{table}

We measured total flux densities at 327 MHz from the new 
images of the observed galaxies. The values obtained are 
reported in Table \ref{sed}.  
\begin{figure*}
\centering
\begin{tabular}{ccc}
\includegraphics[height=5 cm, bb=71 185 519 608]{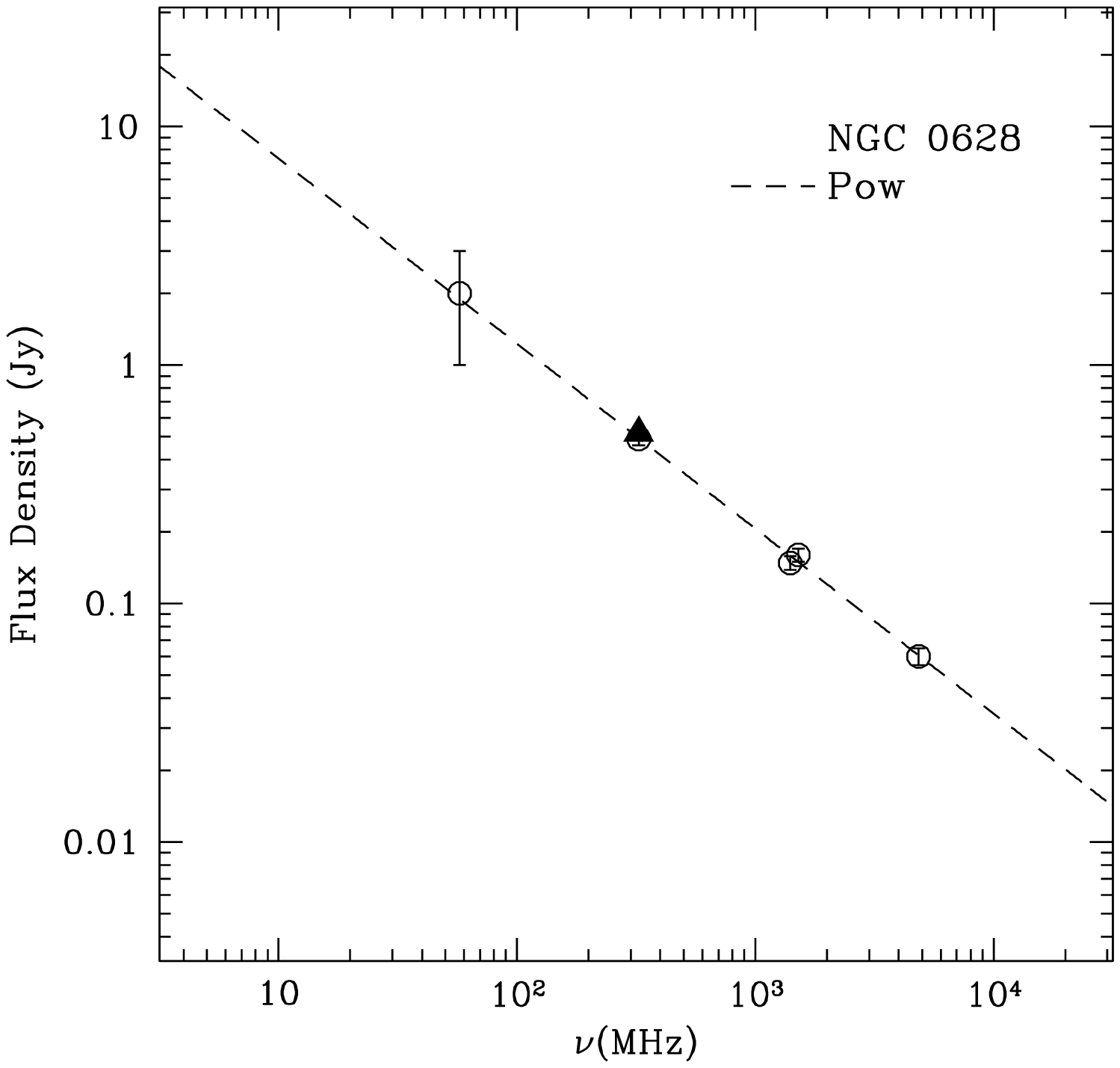}&
\includegraphics[height=5 cm, bb=71 185 519 608]{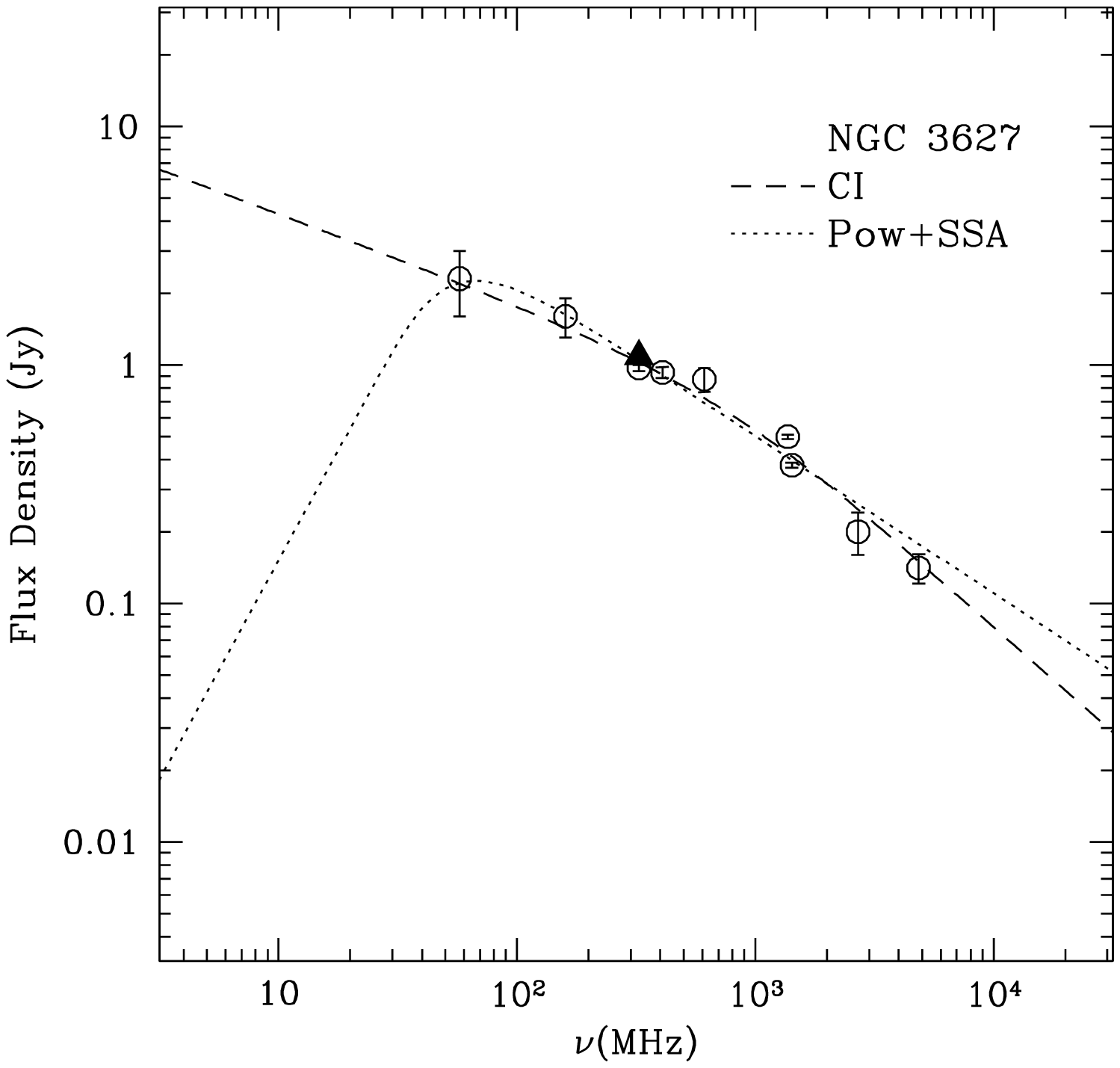}&
\includegraphics[height=5 cm, bb=71 185 519 608]{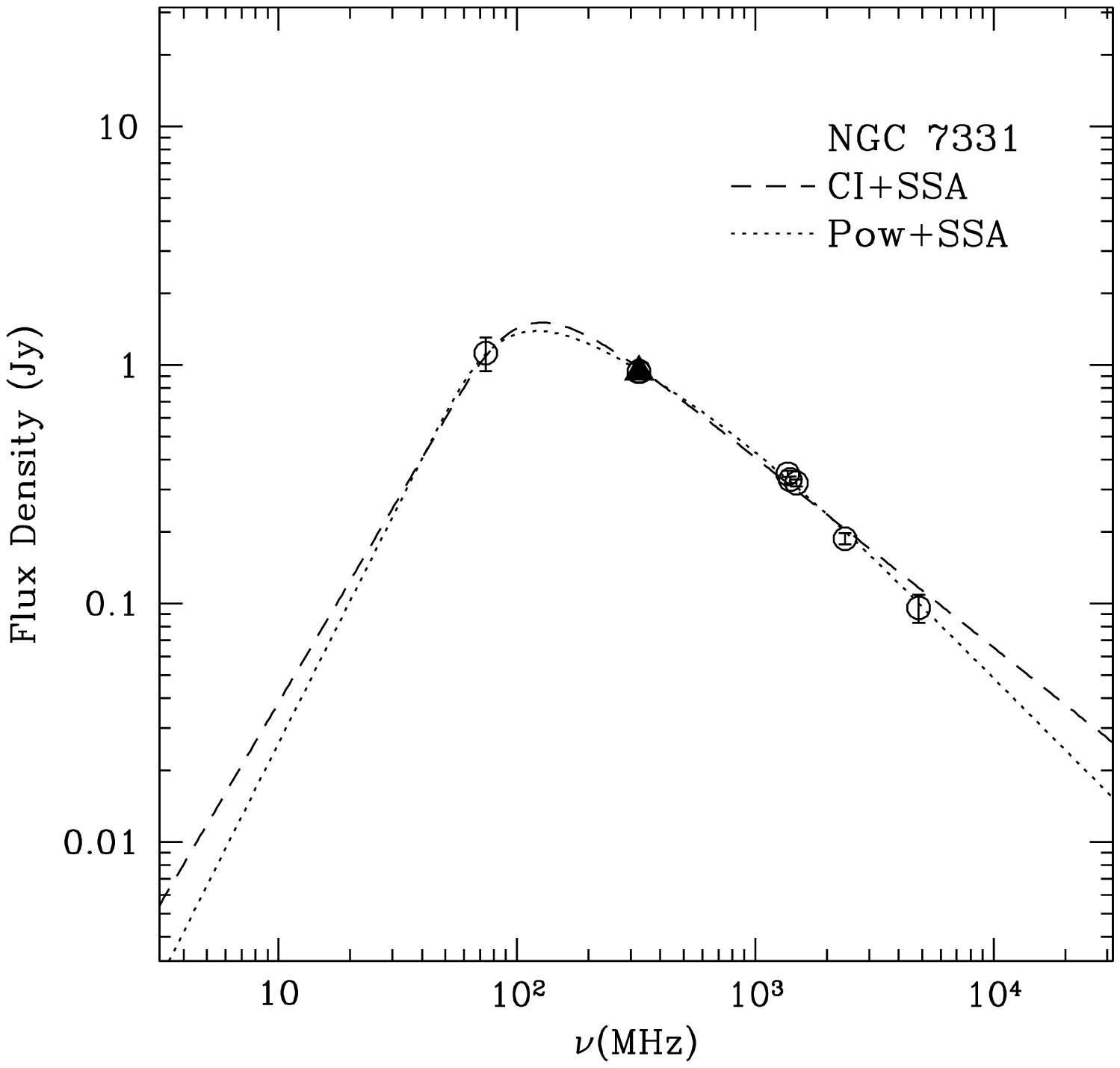}
\end{tabular}
\caption[]{Spectral energy distribution of the observed spiral galaxies.}

\label{Spectr_NED}
\end{figure*}  

\begin{table}[h!]
\caption {Results of the fit with models in detail in Sect. \ref{intspec}:   
\newline
Pow = Powerlaw
\newline
SSA = Synchrotron Self Absorption 
\newline
CI = Continuous injection
}
\begin{center}
 \begin{tabular}{lccccc} \hline \hline
\multicolumn{1}{l}{Source} &
\multicolumn{1}{c}{Model} &
\multicolumn{4}{c}{fitted parameters} \\ 
&& $\alpha$ &$\nu_{\rm br}$ & $\nu_{\rm peak}$ & $\chi^2/\rm ndf$ \\
&& & MHz & MHz&\\
\noalign{\smallskip}
\hline
NGC\,0628 & Pow & 0.78 & --&-- & 0.76 \\
\noalign{\smallskip}
\hline
NGC\,3627 & Pow+SSA &0.66 &-- & 54 & 4.3 \\ 
 & CI & 0.4 & 937 &-- & 2.9 \\
\noalign{\smallskip}
\hline
NGC\,7331 & Pow+SSA &0.79 &-- & 112 & 2.5 \\ 
 & CI+SSA & 0.5 & 1025 &92 & 0.98 \\

\hline \hline
\end{tabular}
\end{center}
\label{fittab}
\end{table}

In order to investigate if some absorption process
 affects the emission from these galaxies, in particular in the range of frequencies 
between 327 MHz and 1.4 GHz, we have analyzed their integrated 
radio spectra.
From a careful assessment of published values between 57 and 5000 MHz 
we derived the integrated radio spectra of the three observed galaxies.
The values of the integrated flux density are reported in Table \ref{sed}. 
The telescopes from which the observations have been taken are reported in correspondence of 
each reference.

Most of the measurements were tied to the flux scale of \cite{baars77}.
 Some values (reported with an indication on  Table \ref{sed}) are on the scale of \cite{wyllie69},
 which \cite{baars77} concluded is very close to their own.

 Figure \ref{Spectr_NED}  shows the integrated radio spectra of NGC\,0628, 
NGC\,3627, and NGC\,7331, in which our new 327 MHz flux density measurements   
are indicated by filled triangles. 
The integrated spectrum of NGC\,0628 is shown on the left panel of Figure \ref{Spectr_NED}.
The data are very well fitted by a single power law with a spectral index of
$\alpha$=0.78.
This value is not unusual for a spiral galaxy; the average
spectral index of spiral galaxies is 0.74 $\pm$ 0.03 \citep{gioia82}.

\begin{figure*}[]
\centering
\begin{tabular}{ccc}
\includegraphics[height=5 cm, bb=57 186 491 613]{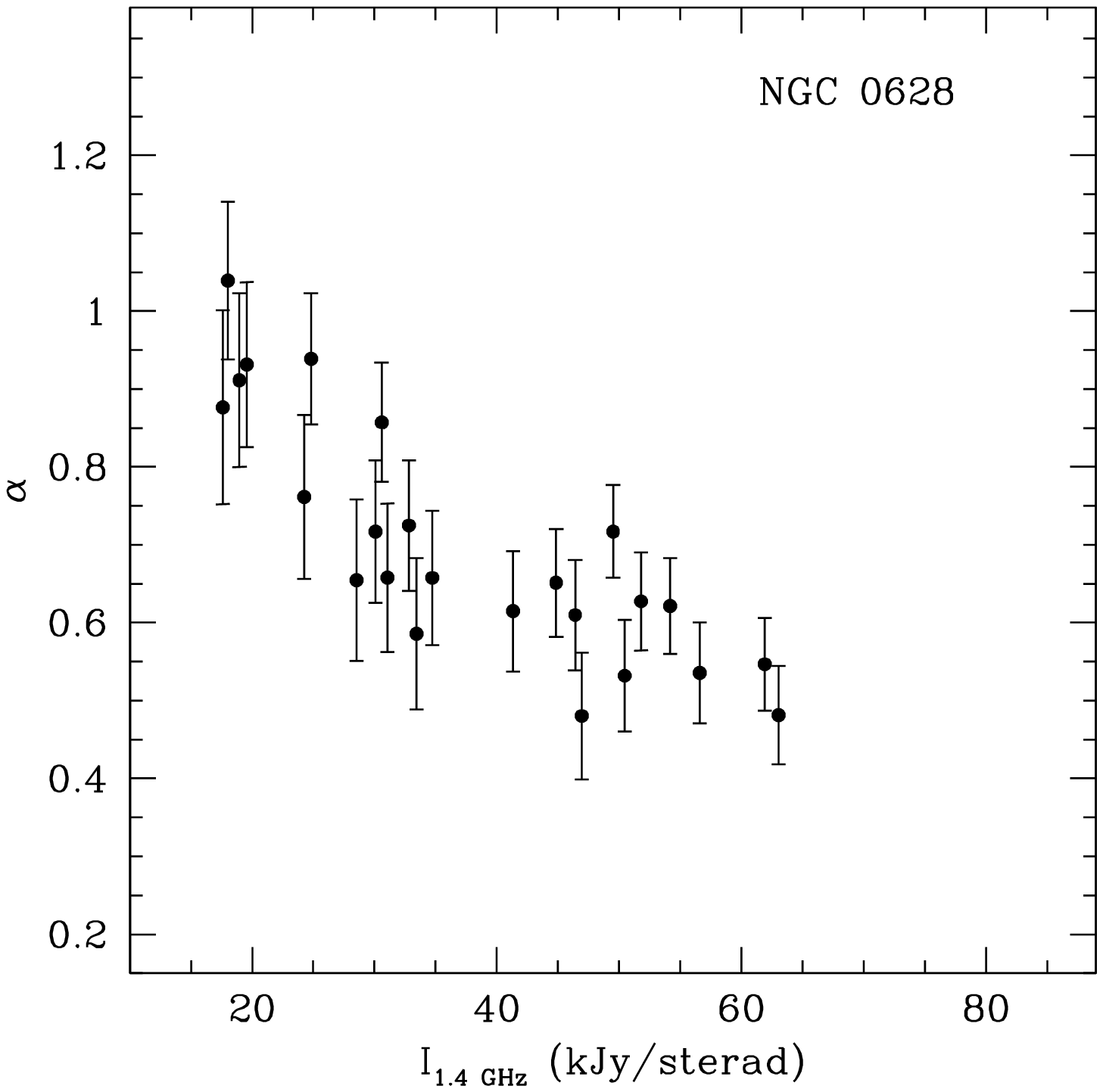}&
\includegraphics[height=5 cm, bb=57 186 491 613]{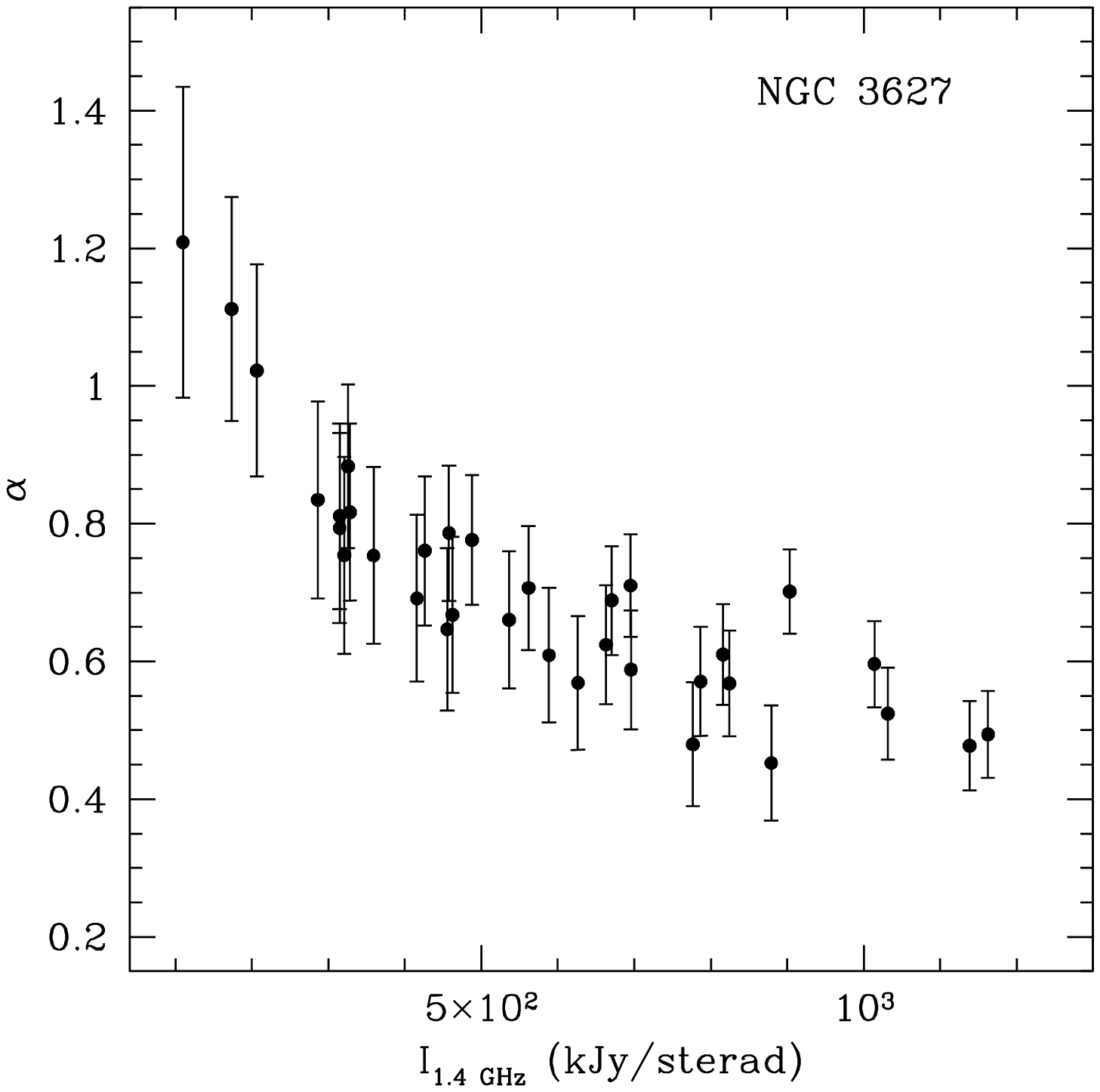}&
\includegraphics[height=5 cm, bb=57 186 491 613]{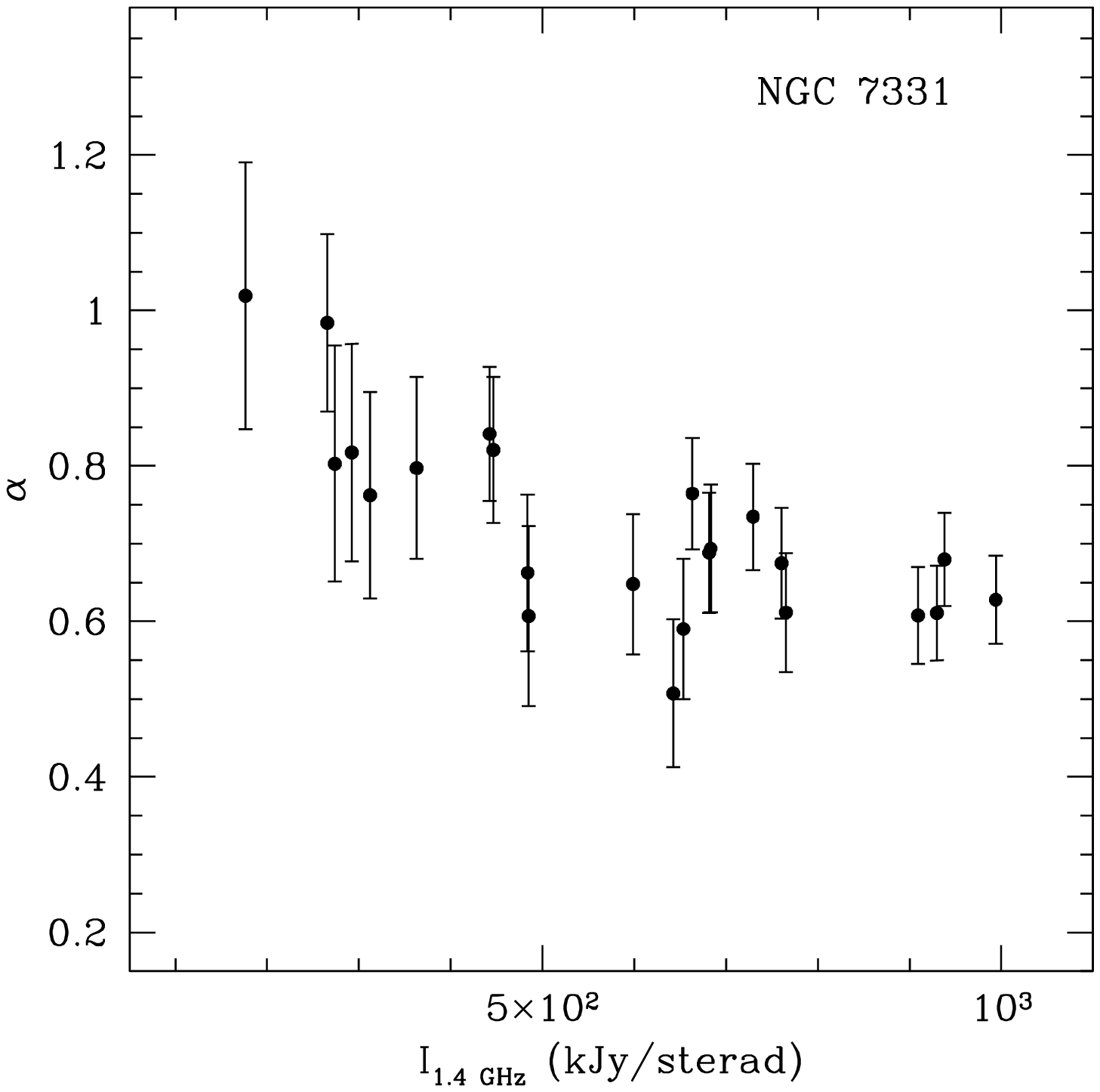}
\end{tabular}
\caption[]{Radio spectral index variation with 1.4 GHz brightness for the three studied galaxies.}
\label{alfaRC}
\end{figure*}

\begin{figure*}[h!]
\begin{center}
\includegraphics[width=0.96\textwidth, bb=1 542 590 840]{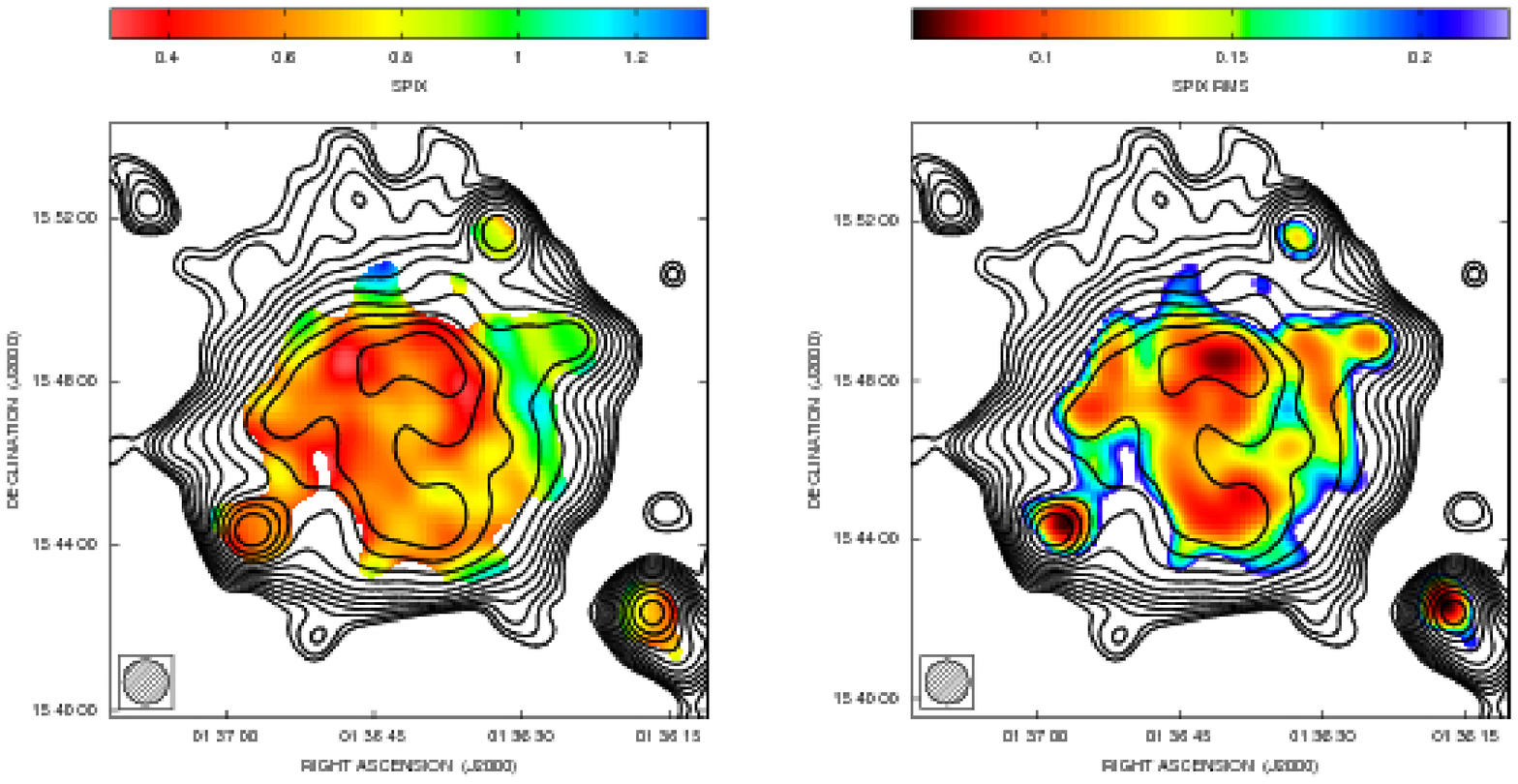}
\end{center}
\caption[]{{\bf{Galaxy NGC\,0628 --}}~{\it Left panel}: The color scale 
represents 
the image of the spectral index 
measured between 327 MHz and 1.4 GHz. Pixels with an intensity level 
above 3 $\sigma$ at both frequencies have been considered.
Contours levels of the radio image at 1.4 GHz, starting from 3 $\sigma$ and 
scaling by a factor of $\sqrt{2}$, are overlaid on the image. 
The FWHM of this image is  66\arcsec$\times$~66\arcsec.
{\it Right panel}: The color scale represents the image of the spectral 
index uncertainty.
}
\label{N0628_spix}

\begin{center}
\includegraphics[width=0.96\textwidth, bb=20 509 530 733]{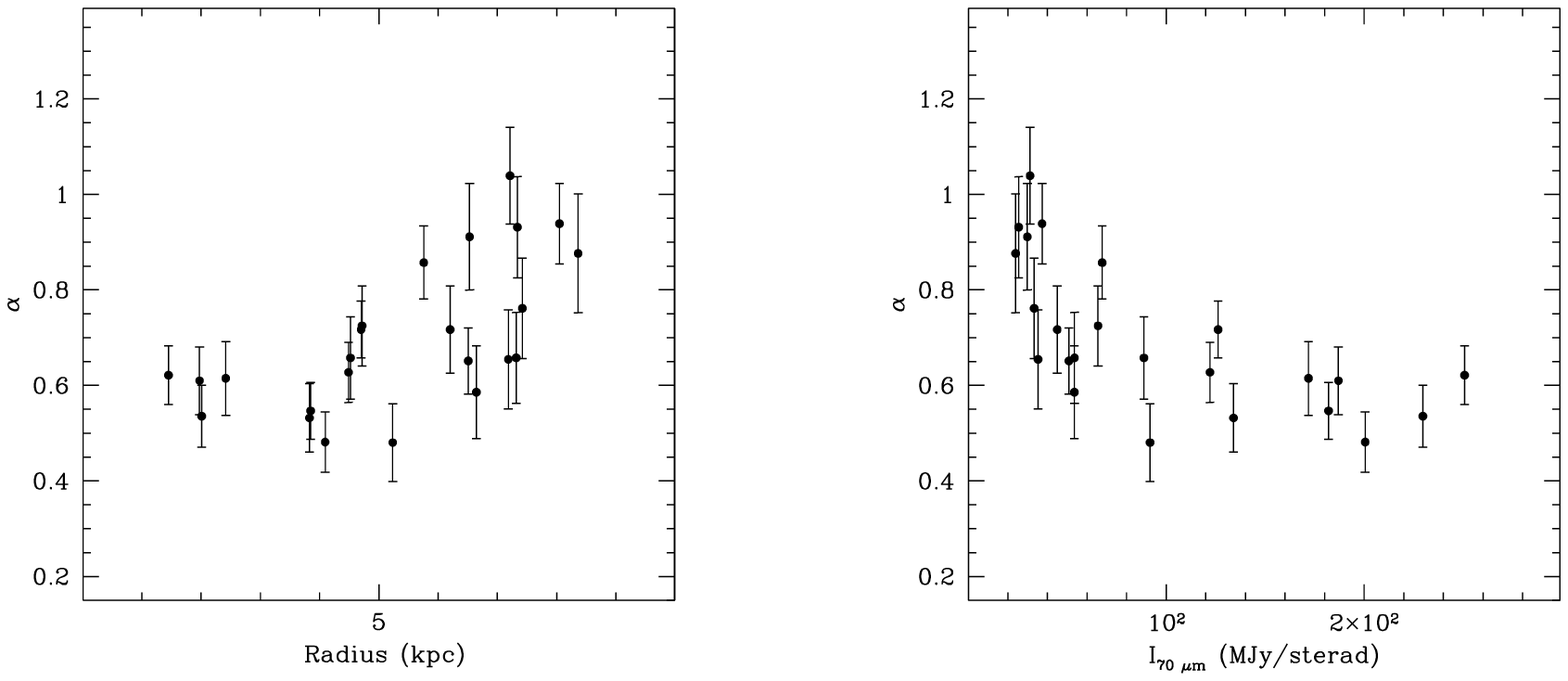}
\end{center}
\caption[]{{\bf{Galaxy NGC\,0628}}: Spectral index radial 
profile (left panel) and spectral index variation with 
IR-70~$\mu$m brightness (right panel).}
\label{N0628_spix_rad}
\end{figure*}

\begin{figure*}[]
\begin{center}
\includegraphics[width=0.96\textwidth, bb=1 542 590 840]{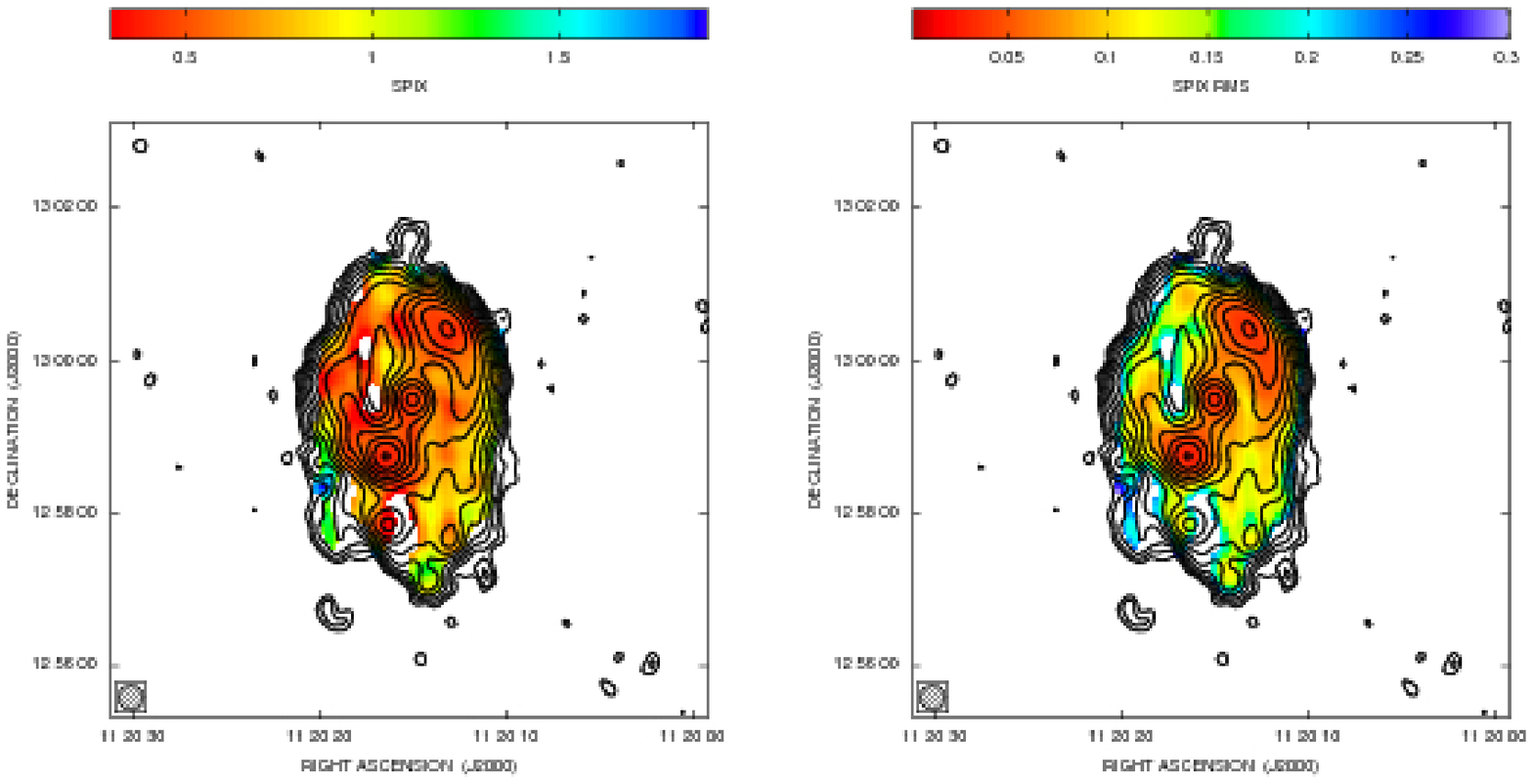}
\end{center}
\caption[]{{\bf{Galaxy NGC\,3627 --}}~{\it Left panel}: The color scale represents 
the image of the spectral index 
measured between 327 MHz and 1.4 GHz. Pixels with an intensity level 
above 3 $\sigma$ at both frequencies have been considered.
Contours levels of the radio image at 1.4 GHz, starting from 3 $\sigma$ and 
scaling by a factor of $\sqrt{2}$, are overlaid on the image. 
The FWHM of this image is  20\arcsec$\times$~20\arcsec.
{\it Right panel}: The color scale represents the image of the spectral 
index uncertainty.
}
\label{N3627_spix}

\begin{center}
\includegraphics[width=0.96\textwidth, bb=20 509 530 733]{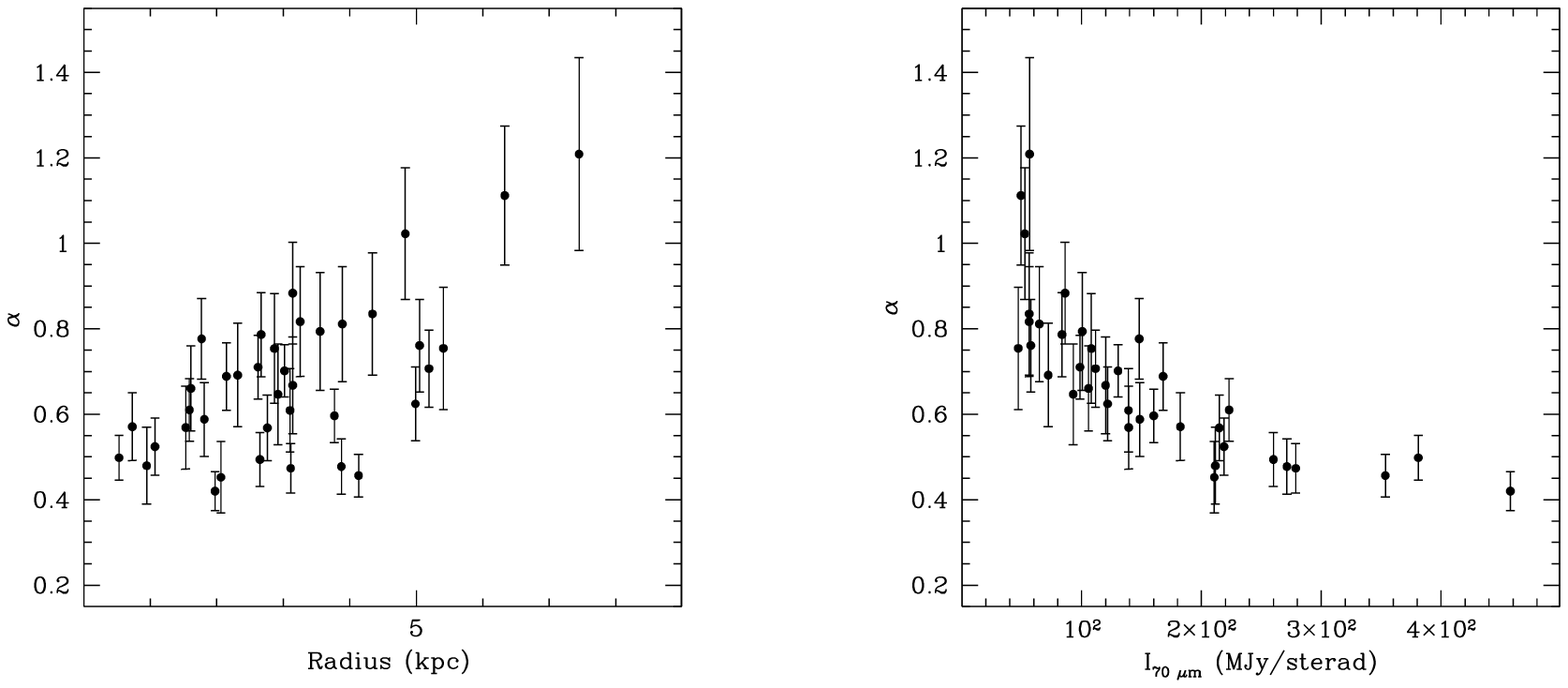}
\end{center}
\caption[]{{\bf{Galaxy NGC\,3627}}: Spectral index radial 
profile (left panel) and spectral index variation with 
IR-70~$\mu$m brightness (right panel).
}
\label{N3627_spix_rad}
\end{figure*}

\begin{figure*}[]
\begin{center}
\includegraphics[width=0.96\textwidth, bb=1 542 590 840 ]{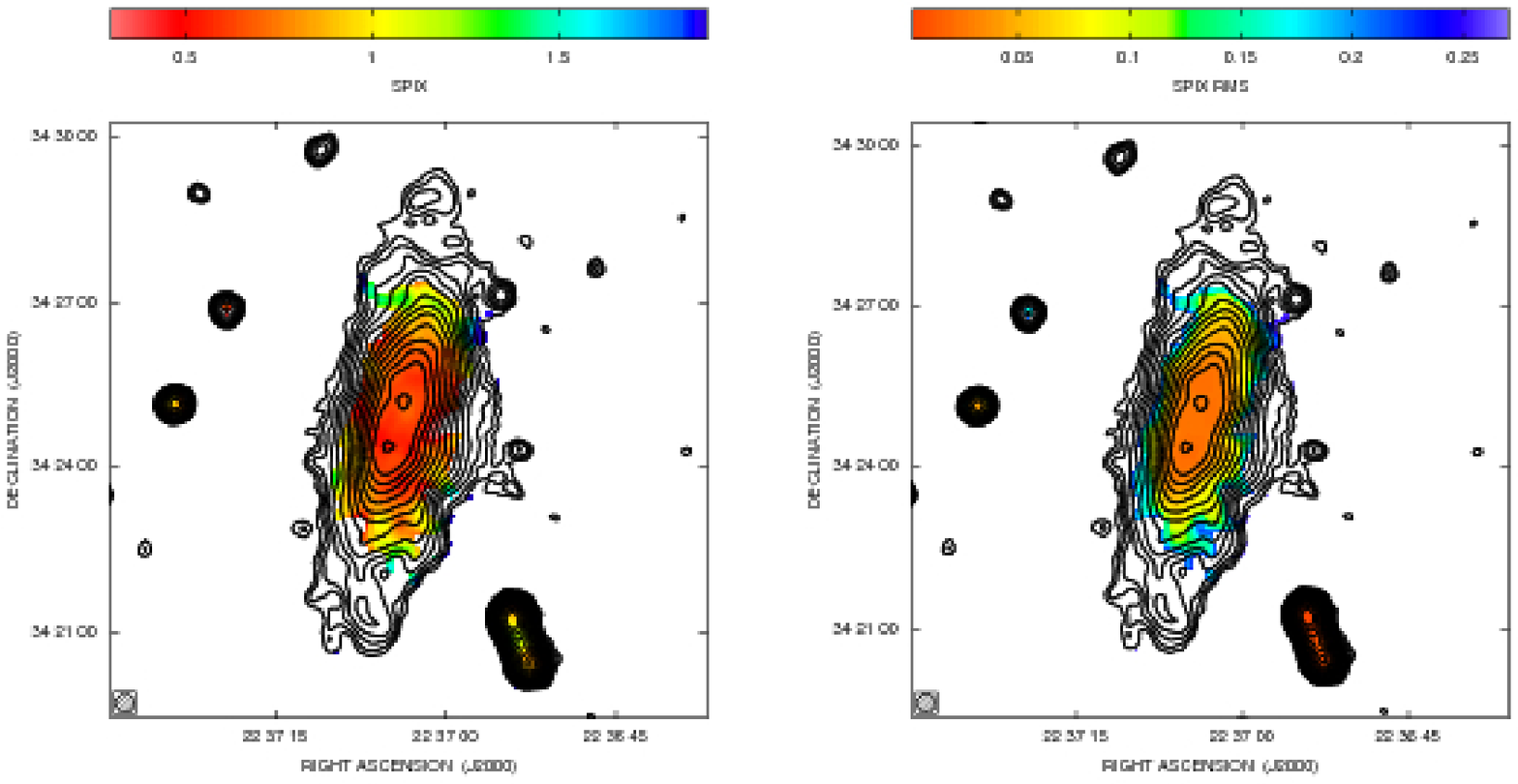}
\end{center}
\caption[]{{\bf{Galaxy NGC\,7331 --}}~{\it Left panel}: The color scale represents 
the image of the spectral index 
measured between 327 MHz and 1.4 GHz. Pixels with an intensity level 
above 3 $\sigma$ at both frequencies have been considered.
Contours levels of the radio image at 1.4 GHz, starting from 3 $\sigma$ and 
scaling by a factor of $\sqrt{2}$, are overlaid on the image. 
The FWHM of this image is  20\arcsec$\times$~20\arcsec.
{\it Right panel}: The color scale represents the image of the spectral 
index uncertainty.
}
\label{N7331_spix}

\begin{center}
\includegraphics[width=0.96\textwidth, bb=20 509 530 733]{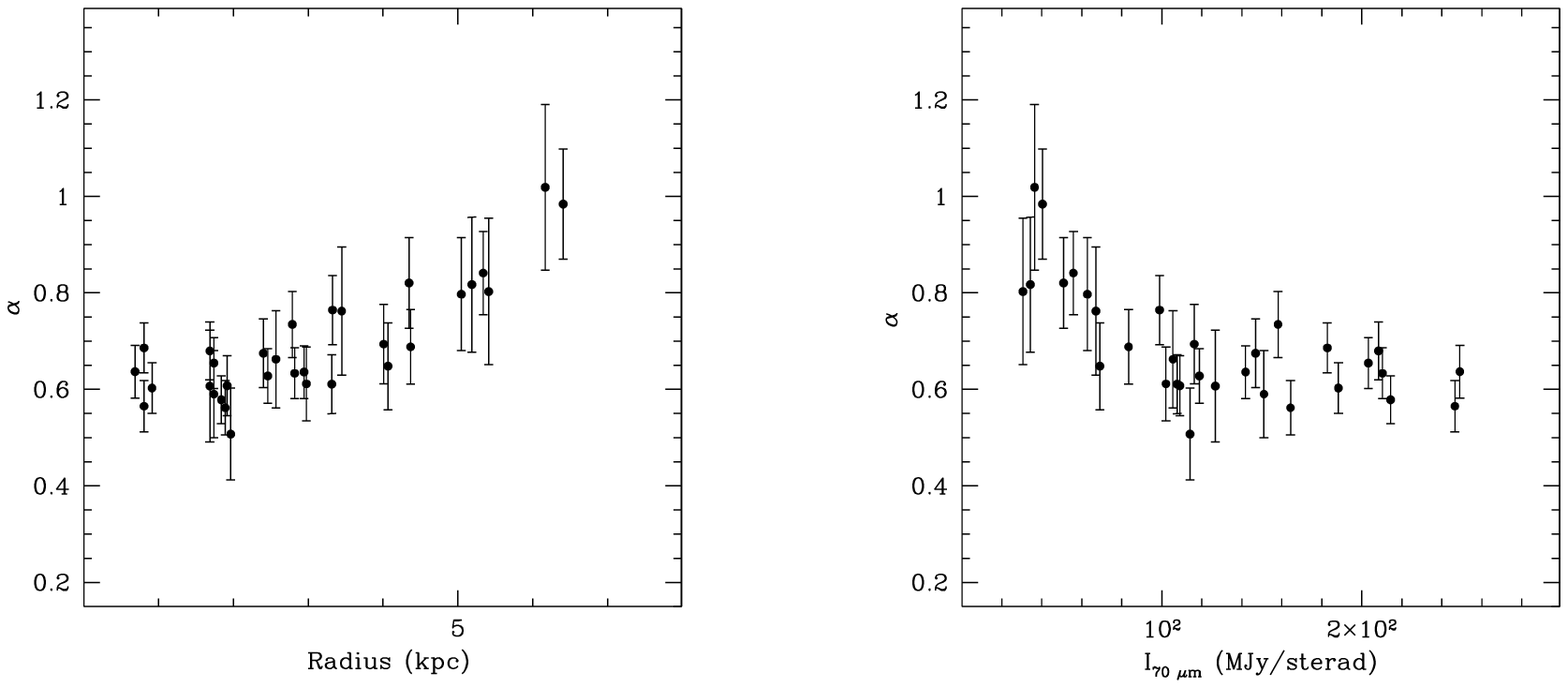}

\end{center}
\caption[]{{\bf{Galaxy NGC\,7331}}: Spectral index radial 
profile (left panel) and spectral index variation with 
IR-70~$\mu$m brightness (right panel).}

\label{N7331_spix_rad}
\end{figure*}

The integrated spectra of galaxies NGC\,3627 and NGC\,7331 
present deviations from the classical power law.
Most spiral galaxies do indeed show a change in spectral index 
with frequency and different possible reasons have been proposed 
(e.g., \citeauthor{israel90} \citeyear{israel90}; \citeauthor{hummel91} 
\citeyear{hummel91}; \citeauthor{pohl92} \citeyear{pohl92}).
We fitted the integrated spectra of these galaxies by introducing an absorption at low frequency.
This seems to be necessary in particular to fit the low frequency point 
in NGC\,7331.
The integrated spectra have been fitted with 
a power law, S$_{\rm Pow}$(${\nu}$) $\propto \nu^{-\alpha}$, modified by low-frequency 
absorption as in \cite{basicsync}:
\begin{equation}
S(\nu)\propto(\nu/\nu_{\rm peak})^{(\alpha+2.5)} 
(1- e^{-(\nu/\nu_{\rm peak})^{-(\alpha+2.5)}})\cdot S_{\rm Pow}(\nu)
\label{pach}
\end{equation}

\noindent
where $\nu_{\rm peak}$ is the frequency at which the optical depth is equal 
to 1, while $\alpha$ is the spectral index in the optically thin 
power law regime.  
For the  optically thick regime we assume a spectral index  2.5, as 
expected in the case of an homogeneous synchrotron self-absorbed source.
The fitted parameters, along with  the reduced 
$\chi^{2}$ values ($\chi^{2}$/ndf, where ndf is the number of degrees of freedom), 
are reported in table \ref{fittab}.

The frequency peaks for both galaxies 
resulting from the fit are at $\nu_{\rm peak} \lesssim$ 100 MHz, thus the region  
between 327 MHz and 1.4 GHz should be relatively free from absorption.

A hint of steepening is observed 
in the spectra above a frequency of $\sim$ 1 GHz. We tried to  
 fit the data with the continuous injection model (CI), described 
in \cite{matteo99}.
 In this model it is assumed that the radio sources are 
continuously replenished by constant flow of fresh relativistic particles
 with a power 
law energy distribution ( N(E)$\propto E^{-\delta}$) and that the magnetic field is constant, 
the radio spectrum has a 
standard shape \cite{kardashev62}, with injection spectral index 
$\alpha_{\rm inj}$=$(\delta - 1)/2 $ below a critical frequency $\nu_{br}$, and 
$\alpha=\alpha_{inj}+0.5$ above $\nu_{br}$. The steepening of the spectral index 
at high frequencies is due to particle energy losses. 


By comparing the $\chi^2/\rm ndf$ values, we 
can notice that the observed high frequency spectra are better fitted by a CI model, rather 
than a power law, for both galaxies. In NGC\,3627 the CI model fit do not requires synchrotron 
self-absorption. Nevertheless, the interpretation of the high frequency break 
should be considered carefully being the assumptions of the CI model too simple 
to describe the complex structures and physics of spiral galaxies. 

In closing,  the analysis of 
integrated spectra of the three galaxies considered in this work  
 shows that the 
region between 327 MHz and 1.4 GHz should be relatively free from absorptions.

\subsection{Local variations in the radio spectral index}
\label{local_spix}

We investigate the variations in the non-thermal spectral index 
as a function of position within the galaxies.
The integrated spectra show that the region between 
327 MHz and 1.4 GHz is relatively free from (self-)absorption, even
in the case of NGC\,7331 where a low frequency turnover seems to be present. Moreover, 
at these frequencies the contribution from thermal 
emission is likely to be minimized, and thus spectral index can be used to study
the spectrum of the non-thermal emission.

The spectral index images between 327 MHz and 
 1.4 GHz, have been obtained by considering only those pixels 
where the brightness was above 3$\sigma$ at both frequencies.
To accurately determine the spectral index distribution 
identical beam size and shape at different frequencies is required.
We convolved NGC\,0628 images to a 66\arcsec~round beam while 
for NGC\,3627 and NGC\,7331 we used a beam of 20\arcsec. 
In addition, to avoid any positional 
offsets, the images were aligned and interpolated to identical  
projections and field center. The primary beam correction has been applied to the 
images at both frequencies.   

The spectral index images, calculated between 327 MHz and 1.4 GHz,
 of the three sources are shown in left panels of 
Figures \ref{N0628_spix}, \ref{N3627_spix} and \ref{N7331_spix}. 
The dynamic range of images at 327 MHz is typically lower than the one at 
1.4 GHz, therefore the cut in these spectral index maps in most of the points is driven by 
the 327 MHz images.
Right panel of the same Figures show the  
spectral index uncertainty.  The comparison of the $\alpha$ image 
with the error image show that most of the observed spectral index features are 
statistically significant.

The 66\arcsec~ beam of NGC\,0628 partly hidden the spiral structure of 
the galaxy, bright clumps in spiral arms in radio images 
correspond to flatter regions in the $\alpha$ image (Fig. \ref{N0628_spix}).
The spectral index image of NGC\,3627 (Fig. \ref{N3627_spix})
clearly shows that the bar and its brights ending regions have a spectral index
flatter than the underlying disk. The radio spectrum of NGC\,7331 (Fig. \ref{N7331_spix}) 
is almost uniformly flat in the galaxy's center and steepens towards the peripheral regions of the disk.

For a quantitative analysis we measured the point-to-point  
brightness on the radio images at 327 MHz and 1.4 GHz as follows.
We overlaid regular grids of rectangular beam-sized boxes 
on the radio images, and we averaged the brightness in each box.
We then calculate the spectral index between 327 MHz and 1.4 GHz. 

Figures \ref{alfaRC}, \ref{N0628_spix_rad}, \ref{N3627_spix_rad} and \ref{N7331_spix_rad}
show the results of this point-to-point comparison. Only points 
exceeding 5 times the rms of the computed spectral index are shown. This 
severe clipping increases our confidence in spectral index variations.

Figure \ref{alfaRC} shows the variations of spectral index with 
the radio brightness in the three studied galaxies.
A common feature in these galaxies is that the spectral index is anticorrelated 
with the radio brightness. Bright regions, like the bar in NGC\,3627 or 
the circumnuclear region in NGC\,7331, are characterized by a flatter 
spectrum with respect to the underlying disk. In all the three galaxies,
the brighter regions have a spectral index of about $\alpha\simeq 0.5-0.6$.
In the faintest regions, the radio spectrum steepen to  $\alpha\simeq 1.0-1.2$.
 
Left panels of  Figures \ref{N0628_spix_rad}, \ref{N3627_spix_rad}, and \ref{N7331_spix_rad}
show the spectral index versus the distance from the center of the 
sources. The anticorrelation between the brightness and the spectral index causes
a systematic steepening of the radio spectrum with the increasing distance from the 
center of these galaxies, where typically brightest regions are located. 

\subsection{Comparison with IR emission}

The far infrared luminosity from galaxies is emitted by interstellar dust 
heated by the general interstellar radiation field and the more intense 
radiation in star formation regions. The 70 $\mu$m emission can be used as star 
formation diagnostic.
To investigate the  connection between star formation sites and 
cosmic rays propagation mechanism, we compare 70 $\mu$m distribution 
to spectral index in galaxies.

We used  70 $\mu$m {\it Spitzer} images \citep{sings03}
convolved to the same beam of our spectral images.
We measured the IR brightness using the same grid described in 
section \ref{local_spix}.
Right panels in  Figures \ref{N0628_spix_rad},\ref{N3627_spix_rad} and \ref{N7331_spix_rad}
show the radio spectral index versus the 70 $\mu$m brightness.
Despite the very different morphologies, the radio spectral index shows
a common behaviour in all the three galaxies: an anticorrelation exists
between the radio spectral index and the infrared brightness.
Regions in which the IR is higher than the average tend to have
a flatter radio spectrum with respect their surroundings.
The radio spectral index shows the same anticorrelation with the radio brightness.
The observed trend is expected on the basis of the local correlation between the 
radio continuum and the infrared emission. This result is consistent with the idea that
the CRe electrons confinement time in the star forming regions is shorter than 
their radiative lifetime.

\section{Summary}
\label{summ}

We have presented new 327 MHz VLA radio images of the three
spiral galaxies NGC\,0628, NGC\,3627, and NGC\,7331. 
Based on these images, 
we measured the integrated flux densities and we 
performed a spatially resolved study of the synchrotron 
spectral index distribution.

We have calculated  the integrated radio spectra of these galaxies,
using the measured integrated fluxes, together with flux 
densities measurements taken from the literature.
In the considered frequency range, the radio spectrum of  NGC\,0628 
is well fitted by a single power law with an index $\alpha\simeq 0.79$.
The integrated spectra of galaxies NGC\,3627 and NGC\,7331 
present deviations from the classical power law.
We fitted these integrated spectra and we 
noticed that the observed high frequency spectra are better reproduced 
by a continuous injection model. 
In NGC\,7331 a synchrotron self-absorption process well reproduce the low frequency 
spectrum.
This is one of the possible
mechanisms producing the low-frequency turnover and further low-frequency observations, are needed 
to confirm it.

We complemented our data set with sensitive archival observations at 
1.4 GHz and 
we studied the variation of the radio spectral
index to within the disks of  these spiral galaxies.
We found that the spectral index is anticorrelated 
with the radio brightness. 
Bright regions, like the bar in NGC 3627 or 
the circumnuclear region in NGC 7331, are characterized 
by a flatter spectrum with respect to the underlying disk. 
This causes a systematic steepening of the spectral index with the increasing 
distance from the 
center of these galaxies. Furthermore,
by comparing the radio images with the 70 $\mu$m images of 
the Spitzer satellite 
we found that a similar anticorrelation exists between the radio 
spectral index and the infrared brightness, 
as expected on the basis of the local correlation 
between the radio continuum and the infrared emission.  
Our results support the idea that in regions of intense 
star formation, where the injection rate of electrons is presumably higher,
the CRe electrons confinement 
time is shorter than their radiative lifetime \citep{matteo05}, 
then the electron diffusion must be efficient.
We also observed the anticorrelation between RC brightness and  
spectral index, this may imply that 
the cosmic ray density and the magnetic field strength are 
significantly higher in these regions than in their surroundings.

Finally, we note that this study adds to a growing list 
of examples confirming the importance of high-resolution, 
low-frequency radio observations in providing 
important tool to understand the physical mechanism  
of production and diffusion of cosmic rays. This also points 
to the great potential of the new generation of 
much more sensitive low frequency instruments, as LOFAR.
Future work will involve a detailed comparison of the radio spectral index 
distribution and the star-formation tracers, as IR and molecular emission, in 
a larger sample of spiral galaxies, in order to analyse the connection of cosmic ray 
production and propagation with the star formation processes.

\begin{acknowledgements}
We would like to thank the anonymous referee for insightful comments which improved 
this manuscript.
E.O.acknowledges financial support of Austrian Science Foundation (FWF) through grant number 
P18523-N16.
The National Radio Astronomy Observatory is
operated by Associated Universities, Inc.,
under contract to the National Science
 Fundation.
This research has made use of the NASA/IPAC Extragalactic Database (NED) which is 
operated by the Jet Propulsion Laboratory, California Institute of Technology, 
under contract with the National Aeronautics and Space Administration
and of CATS database Astrophysical CATalogs support System.

\end{acknowledgements}

\bibliography{paper_LOW}
\bibliographystyle{aa}

\end{document}